\documentclass[12pt]{article}
\usepackage{sidecap}
\usepackage{wrapfig}
\usepackage{graphicx}
\usepackage{threeparttable}
\usepackage[round]{natbib}
\usepackage{times}
\usepackage[margin=15mm]{geometry}
\makeatletter
\renewcommand\section{\@startsection {section}{1}{\z@}%
                       {-1ex \@plus -0.25ex}%
                       {0.25ex}%
                       {\normalfont\large\bfseries}}
\renewcommand\subsection{\@startsection{subsection}{2}{\z@}%
                       {-1ex \@plus -0.25ex}%
                       {0.25ex}%
                       {\normalfont\bfseries}}
\makeatother

\newcommand{\be}{\begin{eqnarray}}
\newcommand{\ee}{\end{eqnarray}}

\begin{document}
\parskip 5pt
\begin{center}
{

\Large\bfseries%
Why Gravitational Wave Science Needs Pulsar Timing Arrays
\\
And Why Pulsar Timing Arrays Need Both Arecibo and the GBT\\

\large\bfseries 
A Response to the NSF-AST Portfolio Review from the NANOGrav Collaboration\footnote{\rm North American Nanohertz Observatory for Gravitational Waves; http://nanograv.org} 
\rm
\vspace {0.12in}

\Large\bfseries%

}
\end{center}
\vspace*{1ex}

Gravitational waves (GWs) are ripples in space-time that are known to exist
 but have not yet been detected directly.  Once they are, a key feature of any viable  theory of gravity will be demonstrated and a new window on the Universe opened.
  GW astronomy was named as one of five key discovery areas in the \textit{New Worlds, New Horizons} Decadal Report.  Pulsar timing probes GW frequencies, and hence source classes, that are inaccessible to any other  detection method and can uniquely constrain the nonlinear nature
of General Relativity. 
 Pulsar timing is therefore a {\it critical capability} with its own discovery space and potential.   Fulfilling this capability
requires the complementary enabling features of both  the Green Bank Telescope (GBT) and the Arecibo Observatory.  Key features of the science and requirements are:
\begin{itemize}
\itemsep -1pt
\item The pulsar timing approach to GW detection uniquely probes  stochastic  backgrounds and continuous-wave signals at nanohertz frequencies along with non-linear bursts from inspiraling 
 black holes.
\item Current limits from pulsar timing arrays (PTAs) have already
produced strong constraints on cosmic strings and are beginning to
constrain scenarios for supermassive black hole evolution \citep{nanograv12}.
\item New spin-stable millisecond pulsars (MSPs) found in large surveys, like those underway at the GBT and Arecibo,  will greatly increase the sensitivity of PTAs to GWs.   The GBT's wide sky coverage is crucial for
finding MSPs in underpopulated regions of the sky (see Fig.~1).
\item The pulsar timing program requires a broad range of frequencies, sensitivity, observing cadences, and sky coverage that only Arecibo and the GBT together can provide.
\item Arecibo and the GBT allow the US to maintain strong leadership in GW science and other, related areas.
\item The Jansky Very Large Array (JVLA) can play a supporting role in the  overall pulsar program but cannot replace the GBT for either pulsar surveys or for long-term, multi-frequency timing with high cadence.
\item The GBT is the only 100~meter class telescope in a radio quiet zone. Other telescopes (Effelsberg, Lovell, Nancay, Parkes, Sardinia) are also
less capable than the GBT in one or more additional aspects.  
\end{itemize}

{\bf Why is pulsar timing a critical capability?}
Methods for GW detection include resonant bars, laser interferometers,
and pulsar timing \citep[e.g.,][]{s10}.
\emph{Pulsar timing probes GW frequencies that are inaccessible to any of those methods}, allowing study of super-massive black hole binaries and stochastic backgrounds from cosmic strings and the earliest stages of the Universe, and was described as ``even more promising'' than other methods of detection in the Decadal Report.  In addition, \emph{pulsar timing is the only way to  detect GW memory} as a discontinous time-domain effect.   Discovery of this effect will prove the fundamental non-linear nature of General Relativity.  
Pulsar timing demonstrated that GWs exist through spectacular,
high-precision timing observations throughout the 1970s and~1980s \citep{tw89,wnt10}, and it is now poised to directly detect GWs.

 {\bf Why are both Arecibo and the GBT necessary?}
 Arecibo and the GBT are the most sensitive facilities in the world for pulsar timing and therefore make NANOGrav a natural leader in the field. As illustrated in Fig.~1, the GBT is essential for fully sampling the 
correlation curve expected due to a stochastic background of GWs. Without the use of the GBT, NANOGrav's time to detection will be delayed by at least several years. In addition, the time to
 detection will be increased by several years
if the pulsars already discovered and those expected to be discovered
by the GBT cannot be timed using the \hbox{GBT}.
\begin{SCfigure}[][!t]
\centering \includegraphics[width=12cm]{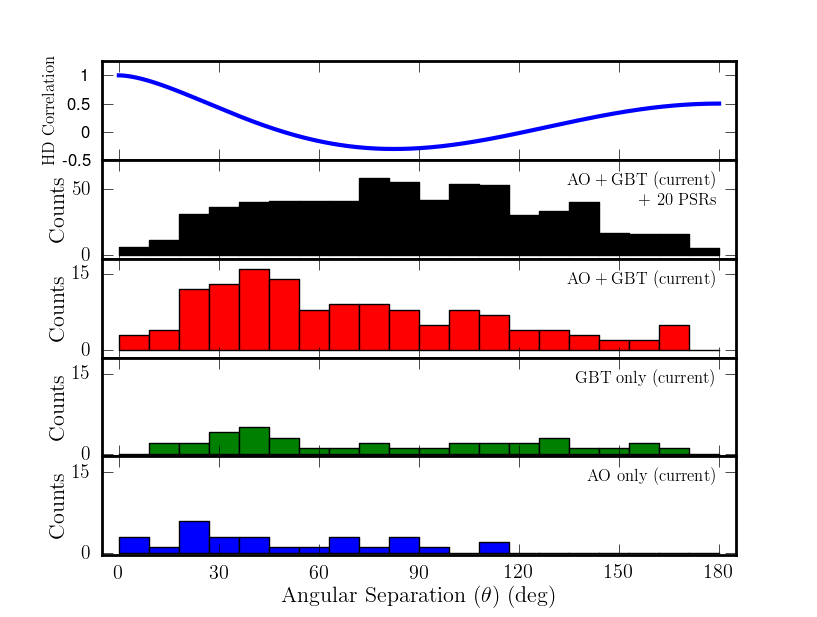}
\caption{ 
\label{fig:detection}
\footnotesize
The top panel shows the  expected correlation in the timing residuals of pulsar pairs as a function of angular separation. This assumes an isotropic stochastic GW background. The other panels show the number
of pairs as a function of  separation for, from bottom to top, MSPs currently timed by NANOGrav with Arecibo (AO), MSPs currently timed by NANOGrav with the GBT, all MSPs currently timed by NANOGrav, and all
 MSPs currently timed plus an additional 20 uniformly distributed MSPs. This plot illustrates the dramatically larger number of pulsar pairs and more complete coverage made possible by the GBT. It also shows the gains possible if we are able to add more MSPs to the array through radio 
pulsar searches in which both the GBT and AO play critical roles. Note the different y-axis scales on the angular correlation histograms.
} 
\end{SCfigure}

{\bf Why can't the JVLA replace the GBT?}
The JVLA can  augment NANOGrav timing but \emph{cannot} replace the \hbox{GBT}.
Firstly,  the GBT is $\sim$50\% more sensitive than the JVLA from 1--2~GHz, where there is an overlap in frequency coverage relevant to pulsar timing. Twice as much time on the JVLA will be required to achieve the same timing precision.  
Secondly, the JVLA provides insufficient  access to radio frequencies less than 1~GHz, needed for mitigating chromatic timing perturbations  due to the interstellar medium, ionosphere, and solar wind. (Three times as much time on the JVLA is needed to achive the same precision on dispersion measure correction).   
In addition, the high cadence needed for pulsar timing is unlikely to be achievable with the JVLA due to its oversubscription.
Finally, a key part of NANOGrav's program is to discover MSPs.   The
GBT is a cornerstone telescope for pulsar surveys due to its
collecting area, low interference environment, and  sky coverage \citep[e.g.,][]{gbt350.i}.
MSP surveys with the JVLA are currently infeasible given the large (5--10~GB/s!) data-rates entailed.

{\bf Why can't other telescopes replace the GBT?}
Several 100~m class telescopes in Europe  combine to form
the Large European Array for Pulsars (LEAP).
Some of these telescopes are not open-access, most are significantly less sensitive and flexible than the \hbox{GBT}, and all have far worse RFI environments. In the case of the open-access telescopes, US astronomers are unlikely to be granted time given the already-entrenched European programs.

The Five-hundred metre Aperture Spherical
Telescope (FAST) and MeerKAT are under construction in China and South Africa, respectively. If they achieve their design specification, they will both be powerful telescopes for pulsar timing.
However, for both telescopes, construction will finish in~2016--2017. It will likely take several
more years for them to approach full capability. The US must 
collaborate with the FAST and MeerKAT projects over the next several years to increase the community of researchers in those countries  and to ensure our access
to these telescopes.
The US will be at a disadvantage in collaborative projects without the use of the \hbox{GBT}.

{\bf What other benefits are there to pulsar timing?}
The NANOGrav program results in high-visibility ancillary science, including mass measurements that constrain the nuclear equation of state, a greater understanding of the formation of compact objects in supernovae, and tests of General Relativity with relativistic binary systems.  Pulsar timing with both Arecibo and the  GBT also provides an important platform for instrumentation development  in partnerships with universities.
  NANOGrav members have also been very successful at involving
students at all levels, from high-school to graduate, in research. The
Pulsar Search Collaboratory and the Arecibo Remote Command Center have
involved over~700 high school students and~100 undergraduates in pulsar searching with the GBT and Arecibo.  Given these considerations, the NANOGrav program is extremely cost effective. Combining the usage fractions of Arecibo and the GBT with community support costs, the ten-year cost of the NANOGrav program is \hbox{\$60M}, a fraction of the cost of LIGO/Advanced \hbox{LIGO}.

\newcommand{\apj}{ApJ}
\newcommand{\apjs}{ApJS}

\end{document}